\documentclass[useAMS,usenatbib]{mn2e}
\usepackage{graphicx}
\usepackage{color}
\usepackage{url}
\usepackage{longtable}

\usepackage [latin1]{inputenc}
\usepackage{appendix}

\definecolor{myred}{rgb}{0.7,0.0,0.2}

\definecolor{myblue}{rgb}{0.0,0.2,0.7}

\definecolor{mygreen}{rgb}{0.2,0.7,0.0}

\title[CV Triples and Winds]{A Speckle-Imaging Search for Close Triple Companions of Cataclysmic Binaries}
\author[M. Shara et al.]{Michael M. Shara$^{1}$\thanks{E-mail: mshara@amnh.org}, Steve B. Howell$^{2}$, Elise Furlan$^{3}$, Crystal L. Gnilka$^{2}$,
\newauthor{Anthony F.J. Moffat $^{4}$  Nicholas J. Scott$^{2}$, and David Zurek$^{1}$}
\\
$^{1}$Department of Astrophysics, American Museum of Natural History, Central Park West at 79th Street, New York, NY 10024, USA\\
$^{2}$NASA Ames Research Center, Moffett Field, CA 94035, USA\\
$^{3}$NASA Exoplanet Science Institute, Caltech/IPAC, Mail Code 100-22, 1200 E. California Blvd., Pasadena, CA 91125, USA\\
$^{4}$D\'epartement de Physique et Centre de Recherche en Astrophysique du Qu\'ebec, Universit\'e de Montr\'eal, Montr\'eal, QC H3C 3J7, Canada
}

\begin{document}

\date{Accepted  Received}

%\pagerange{\pageref{firstpage}--\pageref{lastpage}} \pubyear{}

\maketitle

%\label{firstpage}

\begin{abstract}
The orbital periods of most eclipsing cataclysmic binaries are not undergoing linear secular decreases of order a few parts per billion as expected from simple theory. Instead, they show several parts per million increases and decreases on timescales of years to decades, ascribed to magnetic effects in their donors, triple companions, or both. To directly test the triple companion hypothesis, we carried out a speckle imaging survey of six of the nearest and brightest cataclysmic variables. We found no main sequence companions earlier than spectral types M4V in the separation range $\sim$ 0.02" - 1.2", corresponding to projected linear separations of $\sim$ 2 - 100 AU, and periods of $\sim$ 3 - 1000 years. We conclude that main sequence triple companions to CVs are not very common, but cannot rule out the presence of the faintest M dwarfs or close brown dwarf companions.
\end{abstract}


%==================================================================

\section{Introduction}

Almost all cataclysmic variables (CVs) are comprised of red dwarfs (RD) or red giants transferring hydrogen-rich matter onto the surfaces of their white dwarf (WD) companions \citep{war95}. Mass transfer in these binaries \citep{wal54,kra62,kra64} drives a rich variety of phenomena. The most spectacular of these include the reasonably-well understood accretion-powered outbursts of dwarf novae \citep{osa74,dub18} and the thermonuclear-powered eruptions of classical novae \citep{sst72,pss78}. 

Until the advent of large CCDs and large area synoptic studies, many CVs (other than novae) were found because of their strong emission lines \citep{ste77} or very blue colors \citep{gre86}.  Most CVs are now detected via their outbursts in wide area synoptic surveys \citep{dra14,mro15,mrr15}.

Five to ten erupting Galactic classical novae are usually detected in any given year... often by amateur astronomers. A germinal wide-area, infrared survey of the Milky Way has recently shown that the Galactic nova eruption rate is 46 $\pm$12 novae/yr \citep{dek21}. This confirms that $\sim$ 80-90\% of Galactic novae have been missed over the past century because they are distant and reddened. Deep, high cadence all-sky visible-light surveys, and infrared surveys will greatly increase the outbursting Galactic nova and dwarf nova detection rates in the coming decade. 

GAIA confirms that the closest known of all CVs (WZ Sge) is 43 pc distant , consistent with a space density of CVs that is $\sim$ 5x$10^{-5}$/$pc^{3}$ \citep{pal20}, {\it unless} most CVs ``hibernate" as detached red dwarf-white binaries during most of the time between successive nova eruptions \citep{sha86,hil20}. This GAIA-based space density is $\sim$ four orders of magnitude lower than that of main sequence stars near the Sun. Thus, in zeroth approximation, 1 star in 10,000 near the Sun is currently a mass-transferring CV.

\subsection{CV Orbital periods} 
Other phenomena associated with CVs are much less well understood. As expected for a collection of binaries with randomly oriented orbital planes, $\sim$ 10-15\% of all CVs are deeply eclipsing. Individual eclipsing systems' orbital periods {\it P} can be measured, over a baseline of $\sim$ a decade, with precisions of 1 part in 10$^{7}$. The orbital period distributions of all subtypes of CVs, and their various subclasses, are important tests of evolutionary models of CVs. Population synthesis and binary evolution codes can be coupled with angular momentum loss prescriptions \citep{sch16} to produce predictions of the orbital period distributions of CVs \citep{how01,gol15,sch16}. The models' predictions are poor matches to the observed orbital period distributions of all CVs \citep{sch16}, dwarf novae \citep{kni11} and classical novae \citep{fue21}. In particular, the ratio of short to long period CVs, the observed strong peak of novae in the 3-4 hour period range \citep{fue21} and the shortage of dwarf novae in the 2-4 hour period range \citep{kni11} are seriously discrepant with models' predictions.

\subsection{CVs' {\it dP/dt}}
The rates of change of CV orbital periods {\it dP/dt} are also not well understood. CVs' {\it P} are predicted to be secularly decreasing because the underlying binaries are shedding angular momentum \citep{how01,kni11} via their secondaries' winds \citep{ver81} and gravitational radiation \citep{pac81}. CVs born with P $\sim$ 10 hours should reach P $\sim$ 3 hours in $\sim$ 1 Gyr \citep{kni11, hil20}. This corresponds to {\it dP/dt} of order -10$^{-11}$ cycles/day, or a few parts per billion (ppb)/yr. Contrary to expectations, CVs {\it P} that have been monitored for decades show orbital period decreases, increases and cyclical variations. \citet{bor08} lists 14 well-known eclipsing CVs which display orbital period modulations with amplitudes as large as 7.9 parts per {\it million} (ppm) on timescales of 5 to 35 years. Updates and new additions to this list are given by \citet{pil12}, \citet{boy12}, \citet{bru14}, \citet{pil18}, \citet{pat18} and \citep{sca20}. \citet{bor08} adds ``...there (are) presently no CVs with well-sampled and precise (O - C) diagrams covering more than a decade of observations that do not show cyclical period changes". The amplitudes and oscillatory frequencies of the {\it dP/dt} of all classes of CVs show no clearcut patterns. 

\subsection{Explaining {\it dP/dt}}
\subsubsection{Starspots}
One possible explanation for the modulations of CVs' periods is magnetic activity cycles and starspots in their hydrogen-rich donors \citep{apl92,liv94}. These can greatly vary mass and angular momentum transfer rates on timescales of years. Tomographic studies provide support for this hypothesis, showing large starspots on the secondaries of BV Cen, V426 Oph and SS Cyg \citep{waa07,wat07,hii16}.

\subsubsection{Triple Companions}
A second (and non-exclusionary) explanation is that CVs may display positive or oscillatory {\it dP/dt} because they are in triple star systems.  The distant companion of a CV with years-to-centuries long orbits would periodically displace the center of mass of the CV to approach and then recede from Earth. The variable light travel time to a CV in a triple or multiple star system would yield ``early" and ``late" eclipses, and variations in {\it dP/dt}. Triple companions at linear separations from CVs' centers of mass of 3 AU, 10 AU, and 30 AU would experience orbital periods of $\sim$ 5, 30, and 160 years, respectively.

Triple companions aren't as far-fetched as they might seem at first glance - there are three triple star systems (and seven binaries) amongst the 21 star systems closest to Earth \citep{hen18}. Even higher order multiples are remarkably common \citep{tok18}. Five of the nine nearest triple or higher multiple star systems (GJ 570C, GJ 663, GJ 644, G 041-014 and LP 771-095) contain tertiary components certain or possibly of spectral type M4V or earlier {\footnote{http://www.astro.gsu.edu/RECONS/TOP100.posted.htm}}, corresponding to our observational detection limits (see below).

Most current WDs in CVs must have been red giants with radii $\ge$ 1-3 AU which engulfed their companions in the past \citep{mey79,pac81,law83}. Current triple companions to CVs that were sufficiently distant from the common envelope binary ($\ge$ 3 AU from the red giant) to have avoided engulfment then would have spiraled out to distances $\sim$ 2-3 times larger due to mass loss from the inner, pre-CV binary. Closer in triple companions could have stayed at $\sim$ the same separation, or even moved closer in if they, too, were engulfed by the expanding atmosphere of the red giant. Theoretical considerations \citep{lei20} and numerical simulations of the complex evolution of triple stars \citep{ham21} make it clear that Post Common Envelope Binaries (PCEB) and CVs' triple companions of all masses (less than the initially most massive star in the triple system) and separations are possible. 

\subsubsection{CV, PCEB and EL CVn triples}

Examples of CVs suggested to be triple on the basis of variable {\it dP/dt} include CH Cyg \citep{sko98}, VY Scl \citep{mar00}, V542 Cyg \citep{tho10}, FS Aur \citep{cha12,cha20}, V2051 Oph \citep{qia15}, U Gem \citep{war88,dai09} and EM Cyg \citep{liu21}. The masses of companions required to produce the observed {\it dP/dt} range from those of giant planets, through brown dwarfs, up to $\sim$ 0.8 M$_\odot$.

Eclipsing PCEB that will become CVs also display large departures from expected mid-eclipse times. \citep{zor13} provided a comprehensive overview of eclipse timing variations in PCEB. They found that a remarkable $\sim$ 90\% of PCEBs exhibit period variations. If these variations are due to tertiary companions, their masses are likely in the planetary or brown dwarf mass range. The best-studied example to date is NN Ser \citep{par10}. \citet{beu10}  found excellent agreement with the observed period changes of NN Ser via two additional bodies with masses several times that of Jupiter superposed on the linear ephemeris of the binary. Remarkably, \citet{mar14} found that the two planet model correctly {\it predicted} a progressive lag in eclipse times of 36 s that set in since 2010 compared to the previous 8 yr of precise times. 

Two PCEB binaries that are particularly instructive are Wolf 1130 and V471 Tau. On the one hand, there is no doubt of the triple nature of the pre-CV Wolf 1130 \citep{mac18}, located just 16.6 pc distant from Earth. Wolf 1130 is comprised of a T8 brown dwarf $\sim$ 3000 AU from, and in orbit around an M subdwarf - ONe white dwarf, which is itself in a 0.4967 day orbit. In contrast, the nearby PCEB, pre-CV binary V471 Tau has also often been cited as a triple star candidate \citep{kun11}. However, recent very sensitive observations have ruled out even a brown dwarf companion \citep{har15} with a high degree of confidence.

The EL CVn eclipsing binaries, which are closely related to PCEBs, contain an A- or F- type star and a very low mass ($\sim$ 0.2 M$_\odot$) pre-Helium white dwarf. The low pre-WD masses mean that EL CVn stars must form from dynamically stable mass transfer when the more massive star of the initial main sequence binary was at the end of its main sequence lifetime, or had just entered the sub-giant branch \citep{che17}, with the orbital period of the progenitor binary $<$ 3 days. Virtually all close main sequence stars with orbital periods $<$ 3 days are known to be the inner binaries of hierarchical triple systems \citep{tok06}.
\citet{lag20} examined five EL CVns in detail and found that all had nearby companions, consistent with K-M dwarfs.

\subsection{Motivation and outline of this study} 
Claims of triple companions to most CVs and PCEBs will require decades to centuries to resolve if the only evidence remains eclipse timings. The recent availability of new, very high resolution imaging facilities capable of detecting companions 5 (10) magnitudes fainter than CV binaries at separations as small as 17 (1000) milliarcsec prompted us to search for such companions amongst the closest and brightest known CVs. These angular separations correspond to linear separations of putative triple companions that are relevant for CV {\it P} changes on interesting timescales (see section 3).

In Section 2 we briefly describe the speckle cameras, targets and observations of this study. In section 3 we show the results of our observations, placing stringent limits on optical companions to six of the brightest and closest known CVs. We briefly summarize our results in section 4.
 
\section{Targets and Observations} In this first, exploratory study we observed six of the closest and best-studied CVs. They and their CV subtypes are: IX Velorum (the closest novalike variable), AE Aquarii (an Intermediate Polar in a ``propeller" binary), U Geminorum (the prototypical U Gem-type dwarf nova, and suggested to have a triple companion \citep{dai09}), SS Cygni (the brightest dwarf nova), V884 Her (one of the closest highly magnetic-WD Polars) and AH Herculis (one of the closest Z Cam-type CVs). In Table 1 we list these six targets, their CV subtypes, orbital periods, GAIA-determined distances, the dates of observation, integration time, the Central Wavelengths of the filters used to observe each target, and lower limits in the blue and red passbands on how much fainter any undetected companions must be at 1.2" from the CVs.

The observations were carried out with the twin speckle cameras Zorro and `Alopeke located at the twin 8-meter Gemini-S and Gemini-N telescopes, at Cerro Pachon, Chile and Mauna Kea, Hawaii, respectively. `Alopeke and Zorro are dual-channel imagers using two electron-multiplying CCDs (EMCCDs) as the detectors. Each provides simultaneous two-color, diffraction-limited optical imaging (FWHM $\sim$ 0.017" at 550 nm) of targets as faint as V $\sim$ 17 over a 6.7" field-of-view. Each camera can also resolve wider companions (0.2"-1.5") with very large magnitude differences (up to 8-10 mags). Each has filter wheels providing bandpass limited observations. See \citet{sco18} for a detailed description of the instruments and their filters, as well as the `Alopeke-Zorro Web pages{\footnote{https://www.gemini.edu/instrumentation/alopeke-zorro}}. Each observation is comprised of multiple sets of 1000 exposures of 60 msec each, as well as calibration observations. The data reductions are described in \citet{hor12} and the final data products are discussed in \citet{how11}.

\section{Results}
The results of all observations are shown in Figure 1. The red and blue curves in each figure correspond to the two filters used in each observation. They measure the contrast limits $\Delta$m achieved in magnitudes below the primary target. The reconstructed images, covering a $\sim$ 1.5"x1.5" region centered on each CV are also shown. In several cases (most notably AE Aqr and IX Vel) the surrounding background appears asymmetric and/or ``ringlike". These ``features" are at the noise limit of the images, and are not real.

The key result of this paper is that no close, main sequence triple companions (earlier than spectral type M4V) to any of the six observed target CVs were detected in the separation range 0.02" - 1.2". Our observations are more sensitive in the redder filter, where very close companions ($\rho < $0.1") within 3-5 magnitudes of the CVs are excluded. Farther out (0.1"$<\rho <$1.2"), companions 5-8 magnitudes fainter than the CVs would have been detected. 

Five of the six observed CVs (IX Vel, AE Aqr, SS Cyg, U Gem and V884 Her) are all located at about the same distance from Earth: 100 $^{+15}_{-10} $ pc. At 100 pc the angular separations 0.02", 0.1" and 0.3" correspond to projected linear separations of 2 AU, 10 AU, and 30 AU, respectively. Triple companions at those distances would experience orbital periods $\sim$ 3, 30, and 160 years, respectively, similar to those noted for putative CV triple companions in the Introduction. 

We conclude by noting that our observations are {\it not} sensitive to brown dwarfs or the faintest M dwarfs; high resolution infrared techniques will be needed to search for such companions to CVs.

\clearpage
\begin{figure}

\vspace{-2 mm}

\includegraphics[width=8 cm]{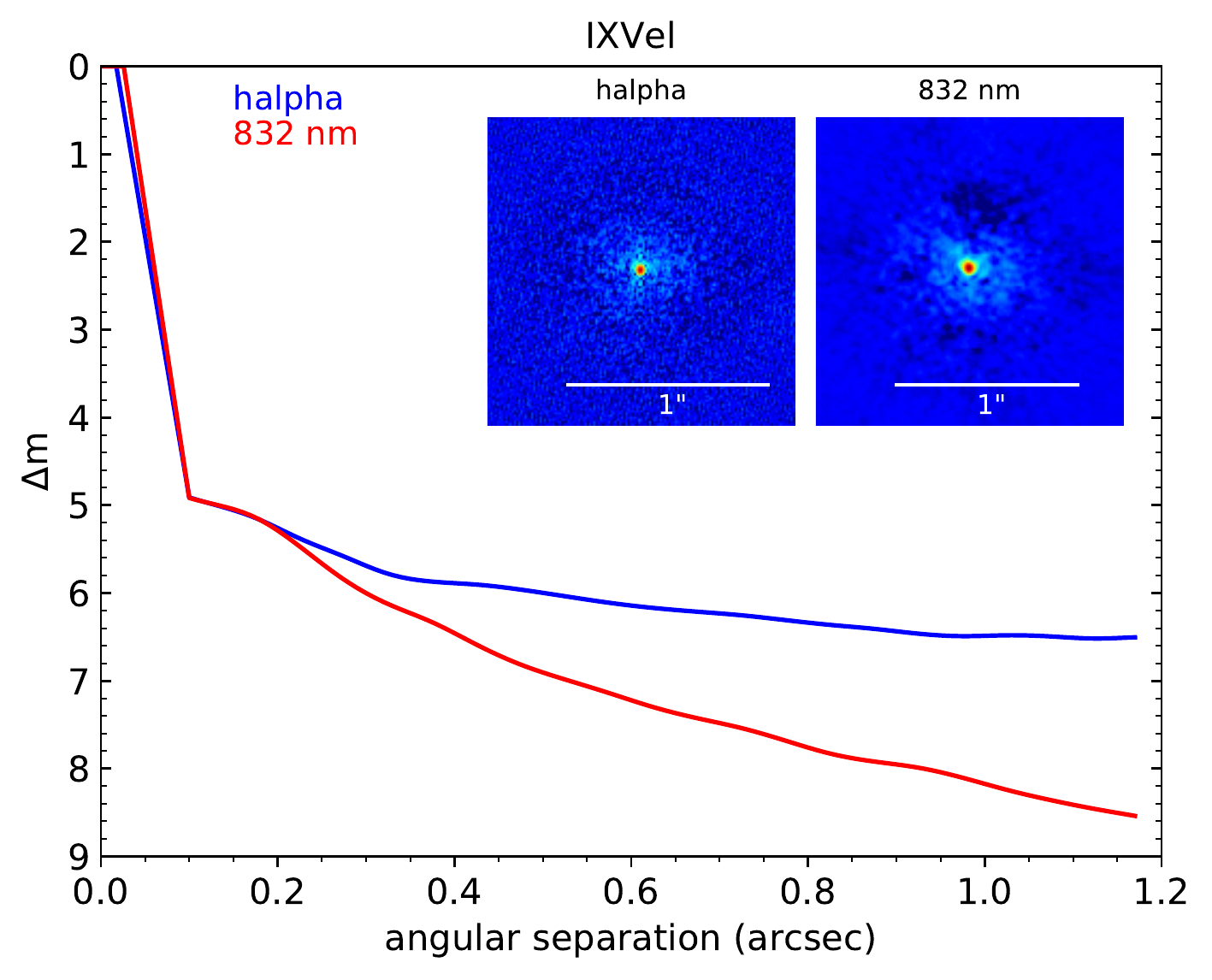}

\vspace{-6.5cm}

\hspace{9 cm}
\includegraphics[width=8 cm]{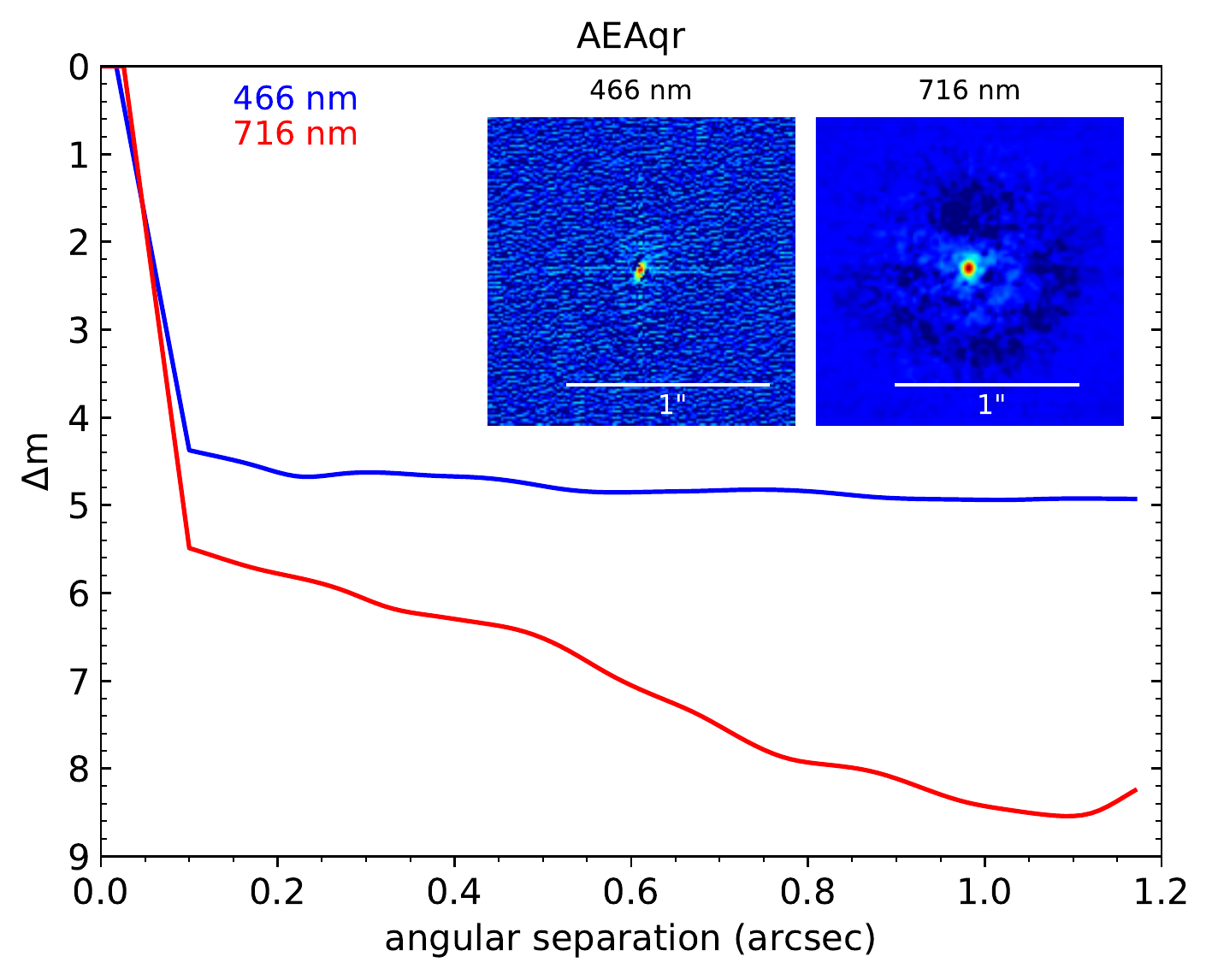}

\vspace{4mm}
\hspace{-1mm}
\includegraphics[width=8 cm]{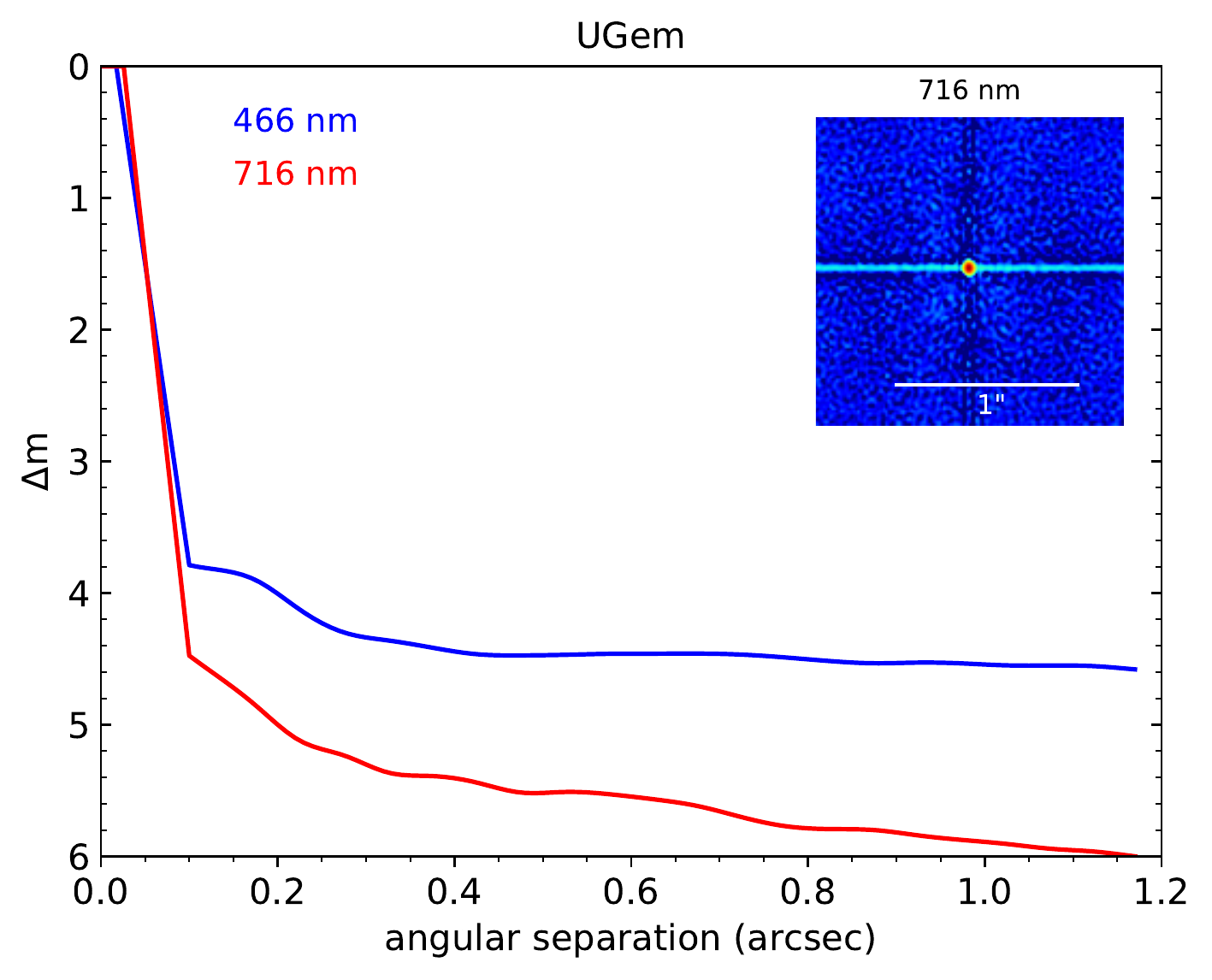}

\vspace{-6.5cm}

\hspace{9cm}
\includegraphics[width=8 cm]{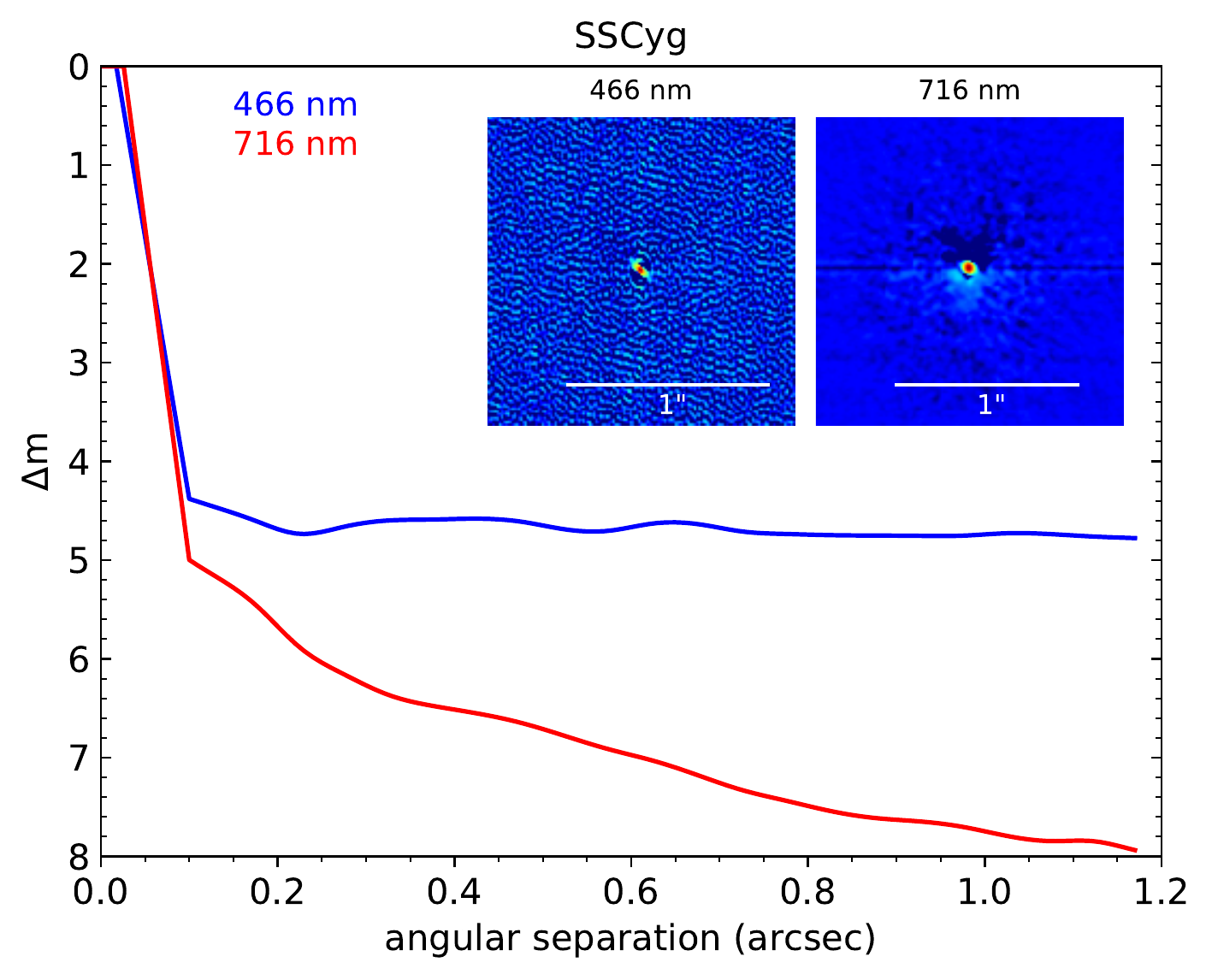}

\vspace{4mm}

\includegraphics[width=8 cm]{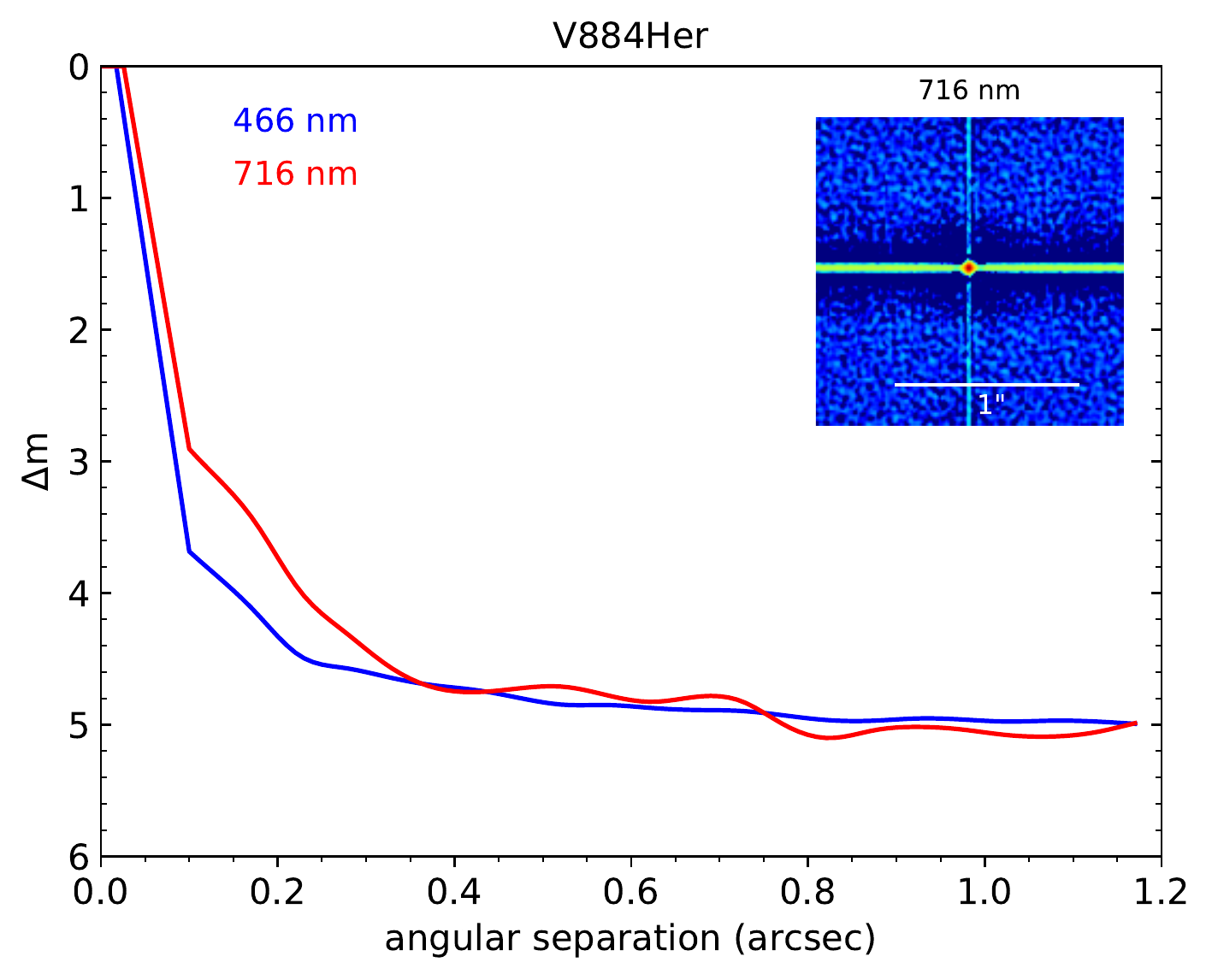}

\vspace{-6.5cm}

\hspace{9 cm}
\includegraphics[width=8 cm]{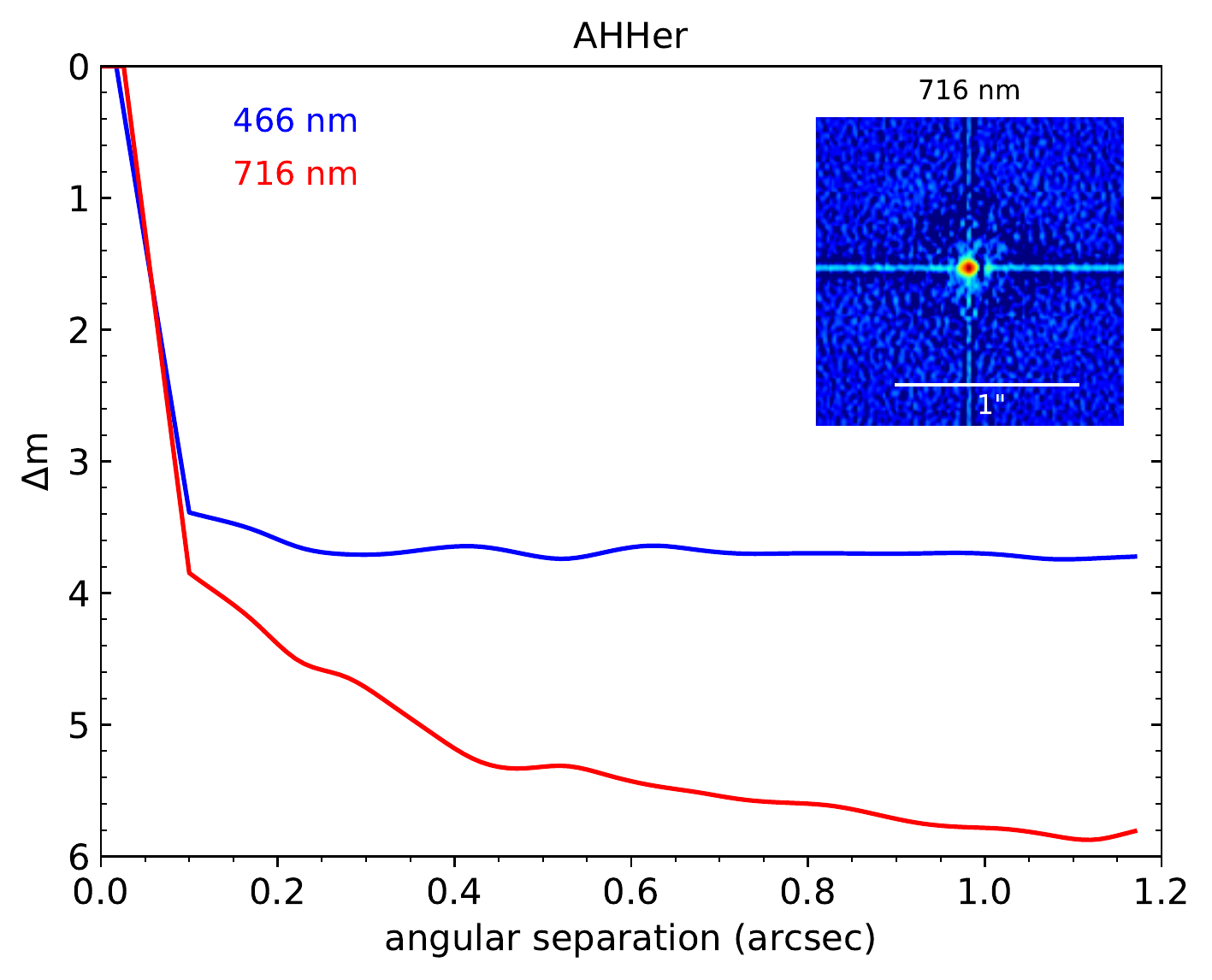}

\caption{The reconstructed images and differential magnitude detection limits $\Delta$m of the six CVs (left to right, top to bottom): IX Vel, AE Aqr, U Gem, SS Cyg, V884 Her and AH Her.  There is no indication of any companion within 100 mas of any of the six stars that is within $\sim$ 4 magnitudes of the CV in either the blue (466 nm) or the red (H$\alpha$ or 716 nm) narrowband filters, respectively. Neither is any companion detected within $\sim$ 4-6 magnitudes (blue filter) or 5-8 magnitudes (red filter) out to 1.2" from any of the CVs. See text for details.}\label{spectra}

\end{figure}

\clearpage

\begin{table*}
 \centering
  \caption{Gemini Speckle Observations of Six CVs}\label{Tid}
  \begin{tabular}{@{}rllllllll@{}}
  \hline
Object &CV Type &$V_{min}$ (mag)& Period(min) &Dist. (pc) & Date Observed&Obs. Time (min) &$\lambda_{filter}$ &$\Delta$mag @ 1"\\
 
\hline

IX Vel &NL & 10.1&279.25 &90.6 $\pm$ 0.2 & 2020-Nov-29 & 30 & 653/832 & 6.5/8.2  \\
AE Aqr & IP &10.4 &592.78 &91.2 $\pm$ 0.5 & 2020-Jun-13 & 20& 466/716 & 5.0/8.4 \\
U Gem & UG & 8.5&254.74 &93.4 $\pm$ 0.3 & 2020-Feb-17 & 35 & 466/716 & 4.5/5.9 \\
SS Cyg & UG  & 8.0&396.19 &114.6 $\pm$ 0.6 & 2020-Jun-15  & 30 & 466/716 & 4.8/7.8 \\
V884 Her & AM &12.8 &113.01 &115.1 $\pm$ 0.3 & 2020-Jun-11  & 8& 466/716 & 4.9/5.1 \\
AH Her & UGZ &10.8 &371.69 &324.3 $\pm$ 3.3 & 2020-Feb-18  & 22& 466/716 & 3.7/5.7  \\

\hline
\end{tabular}

column 2: NL=Novalike Variable; IP=Intermediate Polar; UG=U Geminorum-type dwarf nova; 

AM=AM Herculis CV=Polar; UGZ=Z Camelopardalis-type dwarf nova

column 3: CV's brightest V magnitude

column 8: Filter Central Wavelength (nm)

\end{table*}

\section{Summary and Conclusions} 
Tertiary companions from planets to brown dwarfs to red dwarfs are known or suspected to exist around close binaries. If present, they cause orbital period variations of the inner binary. Such variations are observed for many CVs, though they might be due to other causes. To test whether close stellar companions to CVs might be common, we carried out visible-light speckle observations of six of the nearest/brightest known CVs to try to image nearby triple companions. 

No companions were found in the angular separation range $\sim$ 0.02" - 1.2", corresponding to physical separations of $\sim$ 2 - 100 AU, and orbital periods of triple companions of $\sim$ 3 - 1000 years. While far from definitive, this small survey demonstrates that main sequence triple companions to CVs, of spectral types earlier than M4V, are not common. Our survey is not sensitive to the faintest M dwarfs or brown dwarfs, where high resolution, near-infrared techniques must be employed.

\clearpage
%==================================================================
\section*{Acknowledgments}
The data presented in this paper are based on observations obtained at the international Gemini Observatory, a program of NSF's NOIRLab, which is managed by the Association of Universities for Research in Astronomy (AURA) under a cooperative agreement with the National Science Foundation on behalf of the Gemini Observatory partnership: the National Science Foundation (United States), National Research Council (Canada), Agencia Nacional de Investigación y Desarrollo (Chile), Ministerio de Ciencia, Tecnología e Innovación (Argentina), Ministério da Ciência, Tecnologia, Inovações e Comunicações (Brazil), and Korea Astronomy and Space Science Institute (Republic of Korea). This work was enabled by observations made from the Gemini North telescope, located within the Maunakea Science Reserve and adjacent to the summit of Maunakea. We are grateful for the privilege of observing the Universe from a place that is unique in both its astronomical quality and its cultural significance. Observations in the paper made use of the High-Resolution Imaging instruments Zorro and `Alopeke, which were funded by the NASA Exoplanet Exploration Program and built at the NASA Ames Research Center by Steve B. Howell, Nic Scott, Elliott P. Horch, and Emmett Quigley.  Zorro was mounted on the Gemini South telescope, and `Alopeke on the Gemini-North telescope of the international Gemini Observatory. We thank the Canadian Gemini Time Allocation Committee for excellent feedback and support, and their allocation of telescope time. The observations were obtained under Gemini proposal GN-2020A-Q-110. MS thanks Nathan Leigh and Silvia Toonen for stimulating conversations about the likelihood and origins of triple companions to CVs. We acknowledge helpful suggestions from the referee which improved an earlier version of this paper.
%==================================================================
\section*{Data Availability Statement}

The data underlying this article will be shared on reasonable request to the corresponding author. The raw speckle data are also available on the Gemini archive.

\newpage

\label{lastpage}

\end{document}